# Real-Time Detection of Simulator Sickness in Virtual Reality Games Based on Players' Psychophysiological Data during Gameplay

Jialin Wang

Xi'an Jiaotong-Liverpool University

Hai-Ning Liang[1]

Xi'an Jiaotong-Liverpool University

Diego Vilela Monteiro

Xi'an Jiaotong-Liverpool University

Wenge Xu

Xi'an Jiaotong-Liverpool University

Hao Chen

Xi'an Jiaotong-Liverpool University

Qiwen Chen

Xi'an Jiaotong-Liverpool University

**ABSTRACT**

Virtual Reality (VR) technology has been proliferating in the last decade, especially in the last few years. However, Simulator Sickness (SS) still represents a significant problem for its wider adoption. Currently, the most common way to detect SS is using the Simulator Sickness Questionnaire (SSQ). SSQ is a subjective measurement and is inadequate for real-time applications such as VR games. This research aims to investigate how to use machine learning techniques to detect SS based on in-game characters' and users' physiological data during gameplay in VR games. To achieve this, we designed an experiment to collect such data with three types of games. We trained a Long Short-Term Memory neural network with the dataset eye-tracking and character movement data to detect SS in real-time. Our results indicate that, in VR games, our model is an accurate and efficient way to detect SS in real-time.

**Keywords**: Virtual Reality, Gaming, Simulator Sickness, Machine Learning, EEG, Eye-tracking.

**Index Terms**: [Human-centered Computing]: Human-Computer interaction—Interaction Devices; [Computing Methodologies]: Machine Learning—Machine Learning Approaches

## 1 INTRODUCTION

Virtual Reality (VR) technology has been growing in the last decade, especially in the last few years, with the proliferation of mass-marketed Head-Mounted Displays (HMDs). However, Simulator Sickness (SS) remains a constraint and challenge for VR and has a negative effect on its wider adoption [1][2]. As such, there are significant benefits in finding methods to detect and avoid SS in VR applications, especially in games.

SS is often considered a type of motion sickness which is caused by movement in the environment perceived by the visual system. Currently, there are three possible etiologies of motion sickness: eye movements (EM), sensory conflict (SC), and postural instability (PS) [3]. One of the most significant challenges in SS detection is to quantify it objectively and extract its features.

The most commonly used method to assess SS is the Simulator Sickness Questionnaire (SSQ) [4]. It can be used to quantify SS for activities that could lead to SS symptoms. However, it is exceedingly challenging to quantify real-time SS with SSQ. Although numerous novel SS assessment methods have been proposed to solve this problem [5][6], they require distinct sensors. Such sensors are used to capture Electro Dermal Activity (EDA), Heart Rate (HR), and electroencephalogram (EEG) data. In this research, we proposed a low-cost method to achieve real-time SS detection with current consumer-level VR HMDs with eye trackers (e.g., FOVE, HTC VIVE Pro Eye).

Based on our theoretical analysis and a pilot study, we hypothesize that two in-game features are highly linked to SS: EM and Character Movement (CM). We developed and used our novel own labeling method to break real-time gaming events into classes. Those classes can then be used to train our model and subsequently detect real-time SS in VR. We used a Long Short-Term Memory (LSTM) neural network to train our model using EM and CM. To our knowledge, we are the first to associate these data when looking for SS. Our model can be used to improve the gaming experience whenever SS is detected.

## 2 DATA COLLECTION AND MODEL

### 2.1 Concept and Model

Based on the literature, we pose two hypotheses: $H_1$ CM may cause SS; $H_2$ SS may cause EM. Therefore, EM and CM data may contain patterns that can be associated with SS.

To examine this hypothesis, we created a dataset contained four types of data: (1) magnitude of each eye acceleration, (2) magnitude of character acceleration, (3) speed, and (4) angular acceleration. The magnitude of eye acceleration speed was calculated from 13 types of raw eye movement data recorded from the HTC VIVE Pro Eye. Magnitudes of character acceleration speed and character angular acceleration were calculated using the virtual sensor script in the VR games.

For classification purposes, there are three event IDs in the raw dataset: 0 is normal gameplay without SS; 1 is 6 seconds before SS becomes intolerable; 2 is the pause state (players were requested to pause if sick). We implemented counter-measures to prevent a long blink from being classified as events 1 and 2. Later, we trained two LSTM models, A and B, to detect SS in VR games, prior and after its onset.

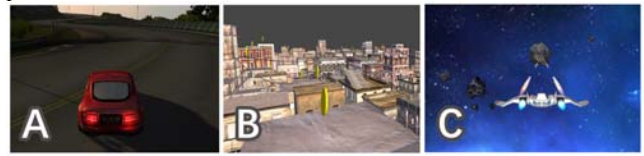

Figure 1: Screenshots of the three VR games used in our experiment for data collection. (A) Racing Car. (B) Parkour. (C) Space Miner.

### 2.2 The Three VR Games

Using Unity3D, we developed three VR games to collect the data during gameplay (see Figure 1). Our pilot run showed that these three VR games could produce enough SS stimulation with different levels during gameplay. To record possible SS symptoms and levels, no SS mitigation techniques have been applied to these VR games.

---

[1] Corresponding author (haining.liang@xjtlu.edu.cn)

In the data collection experiment, Parkour is the only first-person perspective (1PP) VR game among the three games. It can produce twice as many valid SS tags than Racing Car and Space Miner (both 3PP VR games). Prior research also has suggested that 1PP VR games can produce more stimulation than 3PP VR games [7]. Therefore, it is useful to explore 1PP games in more depth to collect data from other 1PP VR games to see if the same findings can be replicated. Further, to improve the performance of pre-SS detection, we need to extract more SS features to find further SS patterns and attempt to test different neural networks. Different length of pre-SS event tags is also worth further exploration due to the difficulty of quantifying SS in milder cases.

## 3 PRELIMINARY RESULTS

Figure 2 shows that Model A has an excellent performance on classifying SS; however, it can only detect post-SS. This model can be used to reduce SS stimulation during the development stage of VR games. It provides developers a low-cost and efficient solution to locate and remove possible SS stimulating aspects during game development or adding ways to mitigate them in the game when compared to existing SS detection methods [8][6].

Our Model B can detect real-time SS events for players who are highly sensitive to SS, but the model is not as sensitive in other cases. Using the model, developers can design pre-defined actions or features in response to SS feedback to improve user experience and gameplay. For example, when SS is detected, the game difficulty can be automatically adjusted to a lower level. There is one crucial difference between Model A and Model B. Model A can only be used to detect post-SS events. Model B can detect part of pre-SS events for players in some games (e.g., 1PP VR games).

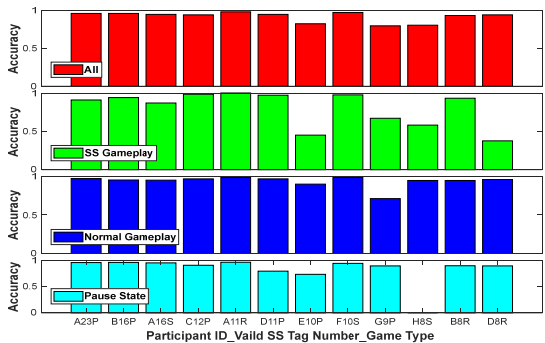

Figure 2: Accuracy of Model A to classify game events.

The application of our real-time SS detection during gameplay is to form a feedback system: it can be used as the basis of an AI-based feedback system in the VR gaming environment. The real-time SS detection model can detect SS from user's real-time data (i.e., eye movement plus character movement data) during gameplay. Then, the VR environment can respond to the SS feedback, possibly reducing SS stimulation in the game.

## 4 LIMITATIONS

Our experiment did not compare its results to more traditional techniques like EDA and HR. Our analysis of EEG data did not find consistent patterns, which could be due to the sample size. Our Model B shows better performance in 1PP VR game (Parkour) which may indicate that the 1PP VR game recordings contains more eye movement features than 3PP VR game recordings. In other words, Model B may only suitable for 1PP VR games.

## 5 CONCLUSION

In this study, we used three different VR games to produce SS stimulation while collecting data. We posed two hypotheses from the etiology of motion sickness and the result of EEG analysis: $H_1$ SS may cause EM; $H_2$ CM may cause SS. The evaluation of our model showed its high performance of post-SS detection for players who are highly sensitive to SS. One of two models, Model A, can be used to detect post-SS for game developers to reduce SS stimulation in their VR games. The other module, Model B, can be used to detect part of pre-SS events for some VR games with strong SS stimulation and players who are highly sensitive to SS. Our preliminary results indicate that our detection method is a effective way to detect SS during gameplay for VR games.